\documentclass[12pt]{article}

\usepackage{overpic}
\usepackage{newtxtext,newtxmath}
\usepackage{graphicx}
\usepackage{caption}
\usepackage[letterpaper,margin=1in]{geometry}
\usepackage{scicite}

\usepackage{url}
\usepackage{xurl}

\usepackage[hidelinks]{hyperref}

\def\scititle{The Rise of AI Search: Implications for Information Markets and Human Judgment at Scale}
\title{\bfseries \boldmath \scititle}
\author{
Sinan Aral,$^{1\ast}$ 
Haiwen Li,$^{1}$
Rui Zuo$^{1}$\and
$^{1}$Massachusetts Institute of Technology, Cambridge, MA, USA.\and
$^\ast$Corresponding author. Email: sinan@mit.edu}
\date{}

\renewenvironment{abstract}{\quotation}{\endquotation}

\makeatletter
\renewcommand{\fnum@figure}{\textbf{Figure \thefigure}}
\renewcommand{\fnum@table}{\textbf{Table \thetable}}
\makeatother

\begin{document}

\maketitle

\begin{abstract} \bfseries \boldmath
We analyzed global, longitudinal AI and traditional search data to spark a conversation about the growth and impact of AI search, and discuss its policy implications for platforms, regulators, scientists and citizens. We executed 24,000 search queries in 243 countries, generating 2.8 million AI and traditional search results in 2024 and 2025. We found a rapid global expansion of AI search and key trends that reflect important, previously hidden, policy decisions by AI companies that impact human exposure to AI search worldwide. From 2024 to 2025, overall exposure to Google AI Overviews (AIO) expanded from 7 to 229 countries, with surprising exclusions like France, Turkey, China and Cuba, which did not receive AI search results. While only 1\% of Covid search queries were answered by AI in 2024, over 66\% of Covid queries were answered by AI in 2025---a 5600\% increase signaling a policy shift on this critical health topic. Our results also show AI search surfaces significantly fewer long tail information sources, lower response variety, and significantly more low credibility and right- and center-leaning information sources, compared to traditional search, impacting the economic incentives to produce new information, market concentration in information production, and human judgment and decision-making at scale. The social and economic implications of these rapid changes in our information ecosystem necessitate a global debate about corporate and governmental policy related to AI search. We hope this discussion and the dataset released with this paper elevate the global dialogue on AI search and its impact.
\end{abstract}

\noindent

\section*{Introduction}

Search is potentially the most consequential application of AI today, simply because we rely so heavily on web search for information about our most significant decisions, from health treatments, to financial decisions, to voting. Google, the world’s most popular search engine, handles 16 billion search queries a day and each of us turns to web search for information four times a day on average, amounting to more than 5 trillion searches a year \cite{axisintelligence2026google}. But the technologies underlying search are shifting, from traditional keyword retrieval and relevance ranking, to AI search. In 2025, OpenAI launched ChatGPT Search to compete with Google's AI Overviews (AIO), Microsoft folded Copilot answers into Bing, and Anthropic, Perplexity and others launched similar AI search products. The result is a decisive shift in how we receive and perceive search information, from navigation (choosing from a list of relevant sources) to synthesis (reading a singular answer delivered in one voice). The implications of this shift for human decision making, the marketplace of ideas and the economics of knowledge production are vast, necessitating a global discussion about policy options to harness the potential of this technology while avoiding its pitfalls.

We have already witnessed significant issues with AI search. For example, it is well known that Large Language Models (LLMs), which power AI search, can produce inaccurate and unsupported information in an effort to create fluent and plausible responses, commonly known as ``hallucinations'' \cite{Huang_2025}. These errors and hallucinations have recently taken center-stage in the AI trust debate as Google’s generative search engine has advised users to ``eat rocks'' for nutrients and to glue cheese onto pizza \cite{eatrock}. After OpenAI launched ChatGPT Search, the Tow Center for Digital Journalism at Columbia University released a study in which it tested the accuracy of ChatGPT Search in referencing quotes from known sources. The study found that out of 200 attempts, ChatGPT Search was ``confidently wrong in 146 cases'' or 73\% of the time \cite{chatgptmisrepresents}. A follow up study in 2025 found that AI search engines cite incorrect news sources 60\% of the time, with Grok demonstrating an astonishingly high 94\% error rate \cite{aisearchhasacitationproblem}.

The Center for an Informed Public at the University of Washington has also noted troubling errors in Google’s AI Search results \cite{searchenginepostchatgpt}. For example, it falsely claimed that studies had shown “women who had an abortion were four times more likely to develop an infection and 2.4 times more likely to experience a postpartum hemorrhage. In some cases, infections can become severe or life-threatening.” In fact, the referenced study found these risks in women who had opted \textit{out} of abortions. When the search string “beneficial effects of nicotine” was typed into Google, the AI results listed “improved mood,” “improved concentration,” “reduced appetite,” and ``reduced muscle tension.” When asked about a fictitious theory called “Jevin's theory of social echoes,” Perplexity AI and the Arc search engine confidently described the made-up theory, complete with graphs and bunk references. In late 2025, it was reported that the hallucination problem is getting worse, not better, as the latest reasoning models hallucinate even more than previous models \cite{aiisgettingmorepowerful}.

\section*{Global Exposure to AI Search}

Given the scale of our reliance on search information, the potential effect of AI search on human decision making, and the potential for errors in AI results, it is critical to understand global exposure to AI search---how often AI appears in search results around the world and trends in our exposure to AI search results by country, topic, and query style over time. Unfortunately, most studies of AI search exposure confound changes in search results with changes in querying behavior, making it impossible to tell if platform policies or our own querying habits are increasing our exposure to AI. To study changes in global exposure to AI search, we focused on Google and executed 24,000 queries, using serpAPI’s Google Search Engine Results API to generate 2.8 million real-world AI and traditional search results in 243 countries. We ran the same identical 12,000 queries in 2024 and again in 2025 to isolate the effects of platform policy, holding querying behavior and style constant. Search queries were randomly sampled from nine data sources containing real user searches, including “Google Natural Questions,” “Most-Searched Google Queries,” “Covid-Related Search Queries,” and “Amazon Shopping Queries” (a full list of query datasets is provided in the SOM). The data show that exposure to AI search results exploded in 2025 and that AI search is now globally pervasive and varies significantly by country, topic and query style.

\begin{figure}[p]
  \centering
  \makebox[\textwidth][c]{
        \includegraphics[width=\textwidth]{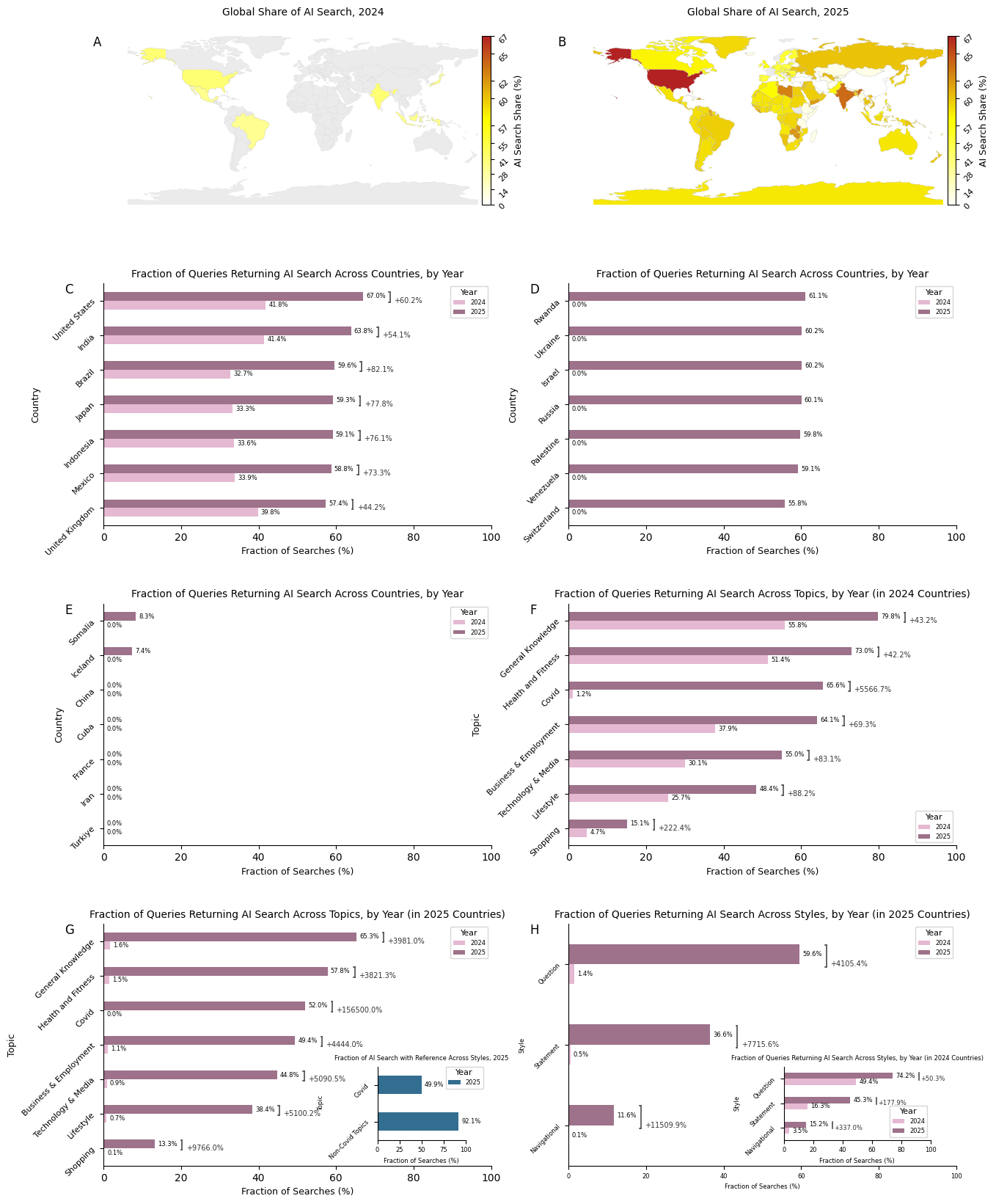}%
        }
  \caption{\protect\phantom{.}} 
  \label{fig:fig0}
\end{figure}


\begin{figure}[t]
  \ContinuedFloat
  \caption{\small{\textbf{Global Exposure to AI Search Results.} This figure shows heatmaps of the fractions of queries that returned AI Search Overviews (AIO) across the world, in 2024 (\textbf{A}) and 2025 (\textbf{B}); the fraction of queries returning AI search results in 2024 and 2025, in the 7 countries exposed to AI search in 2024 (\textbf{C}) (with relative change shown; a “+40\%” indicates a 40\% increase in 2025 compared to the 2024 baseline); in 7 countries exposed to AI search starting in 2025 (\textbf{D}); and in 7 countries with limited to no exposure to AI search in 2024 or 2025 (\textbf{E}) (a full description of these results across all countries is provided in the SOM). (\textbf{F}) shows the fraction of queries returning AI search results across topics in 2024 and 2025 in the 7 countries exposed to AI search in 2024; (\textbf{G}) shows the fraction of queries returning AI search results across topics in 2024 and 2025 in countries exposed to AI search in 2025; (\textbf{G - Inset}) shows the fraction of AI search responses with references across Covid and non-Covid queries; (\textbf{H}) shows the fraction of queries returning AI search results across query style in 2024 and 2025 in countries exposed to AI search in 2025; (\textbf{H - Inset}) shows the fraction of queries returning AI search results across query style in 2024 and 2025 in the 7 countries exposed to AI search in 2024 (with relative changes shown; a “+40\%” indicates a 40\% increase in 2025 compared to the 2024 baseline).}}
\end{figure}

While only 7 countries were exposed to AI search in 2024 (Fig. 1A), by 2025, 229 countries were exposed to AI search to varying degrees (Fig. 1B). The variation in country level exposure is also surprising. For example, while the vast majority of countries exposed to AI search in 2025 saw AI answers between 55\% and 70\% of the time (see Fig 1C and Fig 1D), countries like Somalia and Iceland (which frequently ranks first in overall global internet penetration at over 99\%) saw AI answers only 8.3\% and 7.4\% of the time, respectively (Fig. 1E). Furthermore, while AI search was ``turned on" for 222 new countries in 2025, France, Turkey, Iran, China and Cuba were all excluded from access to AI search (Fig. 1E). In contrast, many other countries, including Ukraine, Russia, Israel, Palestine, Venezuela, Rwanda and Switzerland, went from never having seen AI search answers prior to 2025, to seeing them over 55\% of the time in 2025 (Fig. 1D). AI exposure also increased from 2024 to 2025 in the 7 countries exposed to AI in 2024. Sixty-seven percent (67\%) of the queries in the United States in 2025 were answered by AI, compared to 42\% in 2024 (Fig. 1C). AI search results increased 60\%, 54\%, and 44\% in the U.S., India and the U.K. respectively, while Mexico, Indonesia, Japan and Brazil saw 73\%, 76\%, 78\%, and 82\% increases respectively (Fig. 1C).

The style of the query was extremely influential in determining whether users saw AI results, with questions returning AI answers 60\% of the time, statements 37\% of the time and navigational searches returning AI only 12\% of the time (Fig. 1H). In the seven countries exposed to AI in 2024, all query styles saw increases in exposure to AI search, with questions retrieving AI 74\% of the time (a 50\% increase from 2024), statements 45\% of the time (a 178\% increase) and navigational queries returning AI 15\% of the time (a 337\% increase) (Fig. 1H inset). Logistic regression models indicated that countries explained the most variance in whether searches returned AI results in 2025, followed by the style of queries and then topics (Fig. 2B). As AI search expanded geographically, our logistic regression models revealed that the styles and topics of search queries explained less of the variation in AI search exposure in 2025 than in 2024, while the country of origin of the query explained more, signaling the importance of the geographic expansion of AI search for exposure (Fig. 2A and B).

In 2024, over half of all Health (51\%) and General Knowledge (56\%) queries and only 5\% of Shopping queries and 1\% of Covid queries returned AI results in the 7 countries which were exposed to AI search in 2024 (Fig. 1F) and AI appeared with similar frequencies across those countries (Fig. 1C). By 2025, as AI search expanded to 222 new countries, 65\% of General Knowledge queries, 58\% of Health queries, and 13\% of Shopping queries were answered by AI around the world (Fig. 1G). From 2024 to 2025, in the 7 countries that saw AI answers in 2024, overall exposure to Google AI Overviews (AIO) jumped 67\% on average (Fig. 1C) and increased 42\% for Health queries, 69\% for Business, Finance and Employment queries, 88\% for Lifestyle queries, and 222\% for Shopping queries (Fig. 1F). These topic-level patterns hold when all queries are re-labeled by two alternative LLMs (SOM, Fig. S5). Given that identical queries were implemented across time, the increases, in every topical category and in every country, reflect changes in Google's policies or underlying technology, rather than changes in user querying behavior.

In examining these accelerating topical shifts in AI search exposure, our analysis also uncovered clear policy decisions by AI companies regarding when to expose the public to AI search results on sensitive, regulated topics. For example, in 2024, only 1\% of Covid queries returned AI search results worldwide (Fig. 1F). This policy likely reflected prevailing political sentiment in 2024. During the Biden presidential administration in the U.S., the White House actively intervened with digital platforms to rein in Covid misinformation, and a Supreme Court case, \textit{Murthy v. Missouri}, largely upheld the President's right to set such policy \cite{thesupremecourt}. During this time, AI almost never answered Covid search queries anywhere in the world. However, on the first day of his second term, Donald Trump reversed this policy and signed Executive Order 14149, titled ``Restoring Freedom of Speech and Ending Federal Censorship," which condemned Biden's policy and dismantled government efforts to block digital Covid misinformation \cite{restoringfreedom}. Almost overnight, our exposure to AI search results on Covid queries jumped 5600\%, and in 2025 AI answers to Covid queries went from representing 1\% of answers to 66\% of answers globally (Fig. 1F). Furthermore, the rate at which references and citations were shown in response to different types of queries revealed that while references and citations were shown 92\% of the time in AI answers when answering non-Covid queries, references and citations were only shown 50\% of the time in AI answers when answering queries about Covid. Without exposure analysis like ours, such dramatic choices and changes in corporate AI policy are largely hidden from public scrutiny, in part because such decisions, although incredibly consequential, are rarely publicly announced. Unfortunately, as our study could not measure all such policy changes, decisions related to consequential corporate AI search policies are and will remain hidden without interventions that demand more transparency.

 \begin{figure} 
 	\centering
     \makebox[\textwidth][c]{%
         \includegraphics[width=1.0\textwidth]{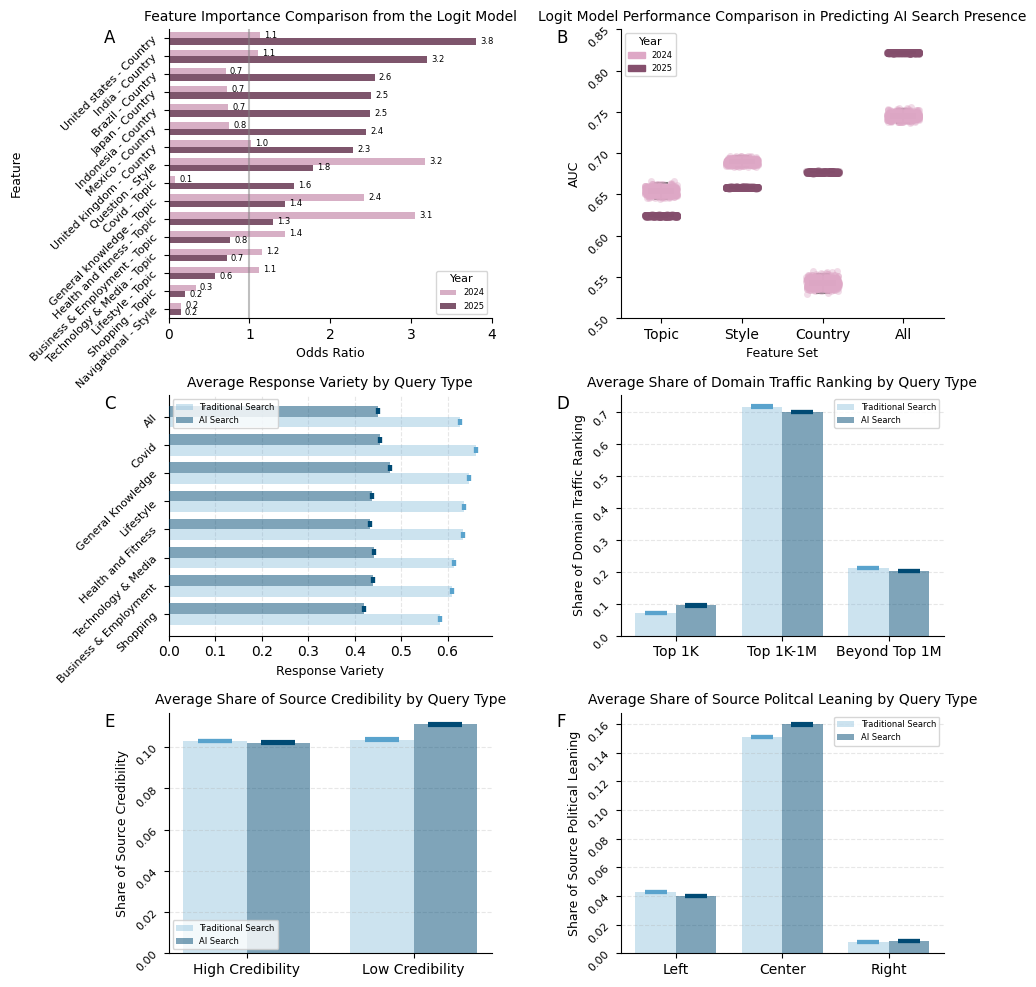}%
    }
         
 	\caption{\small{\textbf{Predictive Features and Information Content of AI Search Results.} This figure displays (\textbf{A}) feature importance comparisons, as odds ratios, from a logistic regression trained on all country, topic and style features, in 2024 and 2025; (\textbf{B}) logistic regression prediction performance comparisons across four feature sets (country, topic, style, all) in 2024 and 2025; (\textbf{C}) the average response variety, measured by aggregated information uniqueness, across traditional and AI search results, by category and for all queries; (\textbf{D}) the average share of domain traffic rankings, measured by cite visits, across traditional and AI search results; (\textbf{E}) the average share of source credibility, measured using the Media Bias / Fact Check database, across traditional and AI search results; and (\textbf{F}) the average share of source political leaning, measured using the Media Bias / Fact Check database, across traditional and AI search results. Additional details of all measurements are provided in the SOM.}}
 	\label{fig:fig2} 
 \end{figure}

\section*{Impacts on Judgment and the Marketplace of Ideas}

Research shows that people respond differently to AI search than they do to traditional search. For example, AI summaries are associated with fewer clicks and shorter sessions. A July 2025 Pew analysis of U.S. search behavior found that users who encountered an AI summary on Google clicked a traditional result in 8\% of visits, versus 15\% when no summary appeared. This represents a near halving of outbound clicks \cite{googleusersareless}. A similar analysis showed that among searches with AIO, the median zero-click rate was 80\% compared with 60\% for searches without AIO \cite{zeroclicksearches}. It is important to note that these studies report associations and are not causal estimates of displacement. However, such dramatic reductions in outbound traffic to sources supplying information to search engines could upend the economic incentives for original reporting, primary research and expert curation. If AI search reduces traffic to Internet sources that feed their existence, we could see dramatic reductions in the incentives to create or report new information. While AI platforms are sending more traffic to publishers than they used to, recent estimates suggest that it is not enough to offset the growth in zero-click sessions driven by summaries that satisfy the query on Google’s pages \cite{ingraphicdetail, agarwal2026impact, Peukert2024AIDisplace, Goel2024AIEating}.

Furthermore, our analysis showed that AI search provided links to URLs in the long tail of the web significantly less than traditional search. AI search referred to the top 1K websites (by traffic) significantly more and from the top 1K to greater than the top 1M websites significantly less than traditional search (Fig. 2D). Such reductions in referral traffic to long tail publishers could lead to market concentration in information production. Understanding whether AI search displaces publisher traffic is critical and more causal research is needed. As LLMs are information synthesizers, not original information producers, if traffic to publishers falls to unsustainable levels, the business models supporting the production of knowledge could be strained or collapse, threatening the health of our entire information ecosystem. If demand for information narrows to only the top publishers, market concentration could threaten the vibrancy and diversity of the marketplace of ideas.\footnote{See the RadioLab Episode titled "What Up Homles?" for an in depth discussion on market concentration and the marketplace of ideas: https://radiolab.org/podcast/what-holmes}

Cognitive changes in how we perceive, evaluate and make decisions based on search information are equally important. In two randomized experiments, LLM-based search tools halved time-on-task and reduced the number of queries, with similar accuracy to traditional search when the model was correct. But they also induced over reliance when the model erred, creating a convenience-accuracy trade-off \cite{effectsofllmbased}. Our exposure to diversity also changes when we rely on AI. Experiments comparing AI search with conventional search engine results pages (SERPs) find that people ask for and consume a narrower set of views when interacting with an LLM that `speaks with one voice' and that an opinionated tone tends to reinforce the user's prior position. In other words, the interface itself can subtly increase selective exposure to information and exacerbate conformation bias, entrenching consumers' original beliefs and furthering polarization \cite{generativeechochamber}. When we analyzed the information content of both AI and traditional search results in the text of the 2.8 million global search results we collected in 2024 and 2025, we found that AI search results exhibited significantly lower response variety than traditional search in every category of information, supporting the idea that AI answers queries `with one voice' (Fig. 2C). This result is robust to differences in writing style and to the choice of embedding model. We paraphrased a stratified 25\% random sample of AI and traditional results with the same paraphrasing model to standardize writing style, and re-encoded all results with a recent state-of-the-art embedding model (SOM, Figs. S3--S4). Our analysis of the URLs linked to in traditional and AI search results also showed that AI search references significantly fewer high credibility information sources and significantly more low credibility information sources (Fig. 2E), and that AI search referenced significantly fewer left leaning sources and significantly more right leaning and centrist sources (Fig. 2F), compared to traditional search.

AI search also affects our trust and behavior in potentially dangerous ways. In a large scale experiment, research found that including reference links and citations in AI search results significantly increased trust in AI search results, even when those links and citations were incorrect or hallucinated \cite{li2025humantrustaisearch}. These results imply that AI designs can increase trust in inaccurate and hallucinated information. Furthermore, references increased trust in AI search results significantly more for people with lower education levels and for those who did not work in technology related industries, which means that less educated and less tech savvy consumers are more vulnerable to misrepresentations and errors in AI search.

Finally, significant concerns exist with attribution, hallucinations, misinformation and errors in AI search results. A 2025 audit of multiple LLMs with web access found that between 50\% and 90\% of response statements were not fully supported—and sometimes contradicted—by the sources cited. Even for a top system, roughly 30\% of individual claims were unsupported and nearly half of the responses lacked complete support, as validated by clinicians \cite{wu2025automated}. When combined with the authoritative nature of AI summaries and our tendency to click less on AI search results, forgoing deeper evaluations, these results suggest that the potential for a meaningful rise in errors of human judgment due to the expansion of AI search worldwide is real.

\section*{Policy Implications of AI Search}

Taken together, these findings suggest that AI search could accelerate decisions and collapse scrutiny; it could re-centralize the web, moving attention from a diversity of sources to a single synthesis; and it could also rewire the economics of the open Internet toward the platform that synthesizes available information into judgments or answers, rather than the sources that produce the knowledge required for those answers. When a handful of AI search engines arbitrate what counts as `the answer' to a given query, there are benefits in speed and accessibility, but there are also risks, including loss of source diversity, increased overconfidence, vulnerability to misinformation, errors of judgment, and structural harm to the economic mechanisms that fund original reporting, research, and expert curation. Crucially, the dramatic transition we are experiencing---from traditional search to AI search---is about path dependence. Once billions of users around the world acclimate to answer-first habits, incentives for publishers, platforms, and policymakers could reorganize around that default. The window to build guardrails---technical, economic, and legal---is now. The right question now is not whether AI search should exist (it will), but how it should operate in a way that preserves speed and access while supporting productive human decision-making and a healthy and sustainable information ecosystem. That requires action from four groups---platforms, regulators, consumers and scientists---with specific, measurable commitments.

\subsection*{Platforms}

Platforms have great power and responsibility in designing the future of AI search with these risks and opportunities in mind. Several tangible and meaningful reforms could go a long way toward enabling the promise and avoiding the peril of AI search. First, platforms like OpenAI, Anthropic and Google should adopt claim-level citations, through which every nontrivial claim in an AI summary would be traceable to a specific passage in a source, with the ability to jump straight to that source with one click. Claim-to-support gaps are common and consequential in AI search \cite{wu2025automated}. Claim-level citations could inspire users to engage in deeper research, countering over reliance on AI summaries, incentivizing and encouraging reflection on a diversity of sources and increasing traffic to information producers and publishers.

Second, platforms should adopt lightweight signals, including confidence highlighting and contested claim badges, that have been shown to reduce over reliance without sacrificing speed \cite{effectsofllmbased}. When reputable sources disagree, this disagreement should be highlighted rather than washed out by AI synthesis. Confidence highlighting and contested claim badges could empower human judgment with full information about the uncertainty of AI answers.

Third, platforms should adopt and adhere to diversity quotas for reliable sources and introduce topic-specific diversity objectives to avoid excessive reliance on a small set of `usual suspects.' Independent audits already show that Reddit, Wikipedia, or YouTube dominate AIO citations \cite{hurtingclick} and it’s time to widen the funnel. Such reforms could not only help nurture information diversity, but could support long-tail publishers most at risk from the economic consequences of the re-centralization of internet traffic that AI search engenders.

Fourth, it is imperative that AI platforms adopt transparent prevalence and impact reporting that includes quarterly statistics, by market and topic, on the percent of queries with AI summaries, their median word count, the median number of sources cited, click-through rates to cited sources, distributions of traffic by site class (e.g. news, gov, edu), and error-rate measurements from internal red-team tests on hallucinations and errors overall and by category and topic \cite{googleusersareless, zeroclicksearches}. As demonstrated by the Covid revelations of our exposure study, such data could help researchers assess the effect of AI search on society and, in turn, help regulators understand the impact of regulations and what additional measures may be needed as AI search expands worldwide.

Finally, platforms should engage in licensing and revenue-sharing to incentivize information production. Paid licensing agreements between platforms and publishers (already emerging in the EU under the AI Act’s transparency regime) can create incentives for a healthy information ecosystem. Such arrangements could rely on standard contracts and machine-readable signals to scale the technology and to connect the impression and engagement measurements directly to payments \cite{euaiact}. Additionally, platforms should regularize and respect “no summary” signals that allow platforms to synthesize content fairly while giving publishers simple automated controls with which to opt out of verbatim summary extraction, while remaining indexable for ranking.

\subsection*{Regulators}

Regulators also need to think clearly about the appropriate legal and policy environment in which AI search serves humanity best. Balancing the impacts on human judgment, misinformation, and the economic structures that support our information ecosystem are essential. Several reforms are worth considering.

First, it is important for governments around the world to demand Answer Engine Transparency (AET) reports. Building on financial stress tests and safety model cards, such reports should require dominant platforms to publish standardized, audited AET metrics, which should include AI search prevalence data (of the type we collected in this paper), topic-by-topic click-through rates, distributions of citations by site class, measured rates of unsupported claims, and complaint volumes by topic. Such metrics could go a long way toward providing the transparency needed to manage the impact of AI search on the world. These reports should be jurisdiction specific and share common methodologies so statistics can be compared across regions. Transparency provisions in the legal frameworks governing digital platforms in the UK and Australia, as well as the EU’s AI Act transparency provisions, already provide legal groundwork through which to mandate and implement such reporting \cite{euaiact,cmatakesfirst}.

Second, appropriate enforcement of licensing and provenance provisions, for example under the EU AI Act, could help prevent the economic hardship AI search could impose on information producers. The EU AI Act already contemplates transparency for general-purpose AI—including training-data summaries and copyright compliance. It is important to clarify, through guidance, the differences between the synthesis, summarization and reproduction of information in AI search and the conditions under which such activity should trigger heightened provenance duties \cite{euaiact}.

Third, it is important to protect the open-web commons. Policymakers should consider a measured “link-out floor” for certain public-interest domains (public health, civic information, emergency response) so that AI answers must visibly surface at least a known minimum number of diverse source links above the fold. Pairing such requirements with fast-track appeal when authoritative sources are systematically under-cited could ease the burden of their implementation.

Finally, we must fund and support independent third party measurement. Governments must fund academic and civil society panels that conduct recurring, open methodology audits of answer prevalence, support quality, and differential impacts on news and information production, analogous to inflation baskets, but for the information ecosystem. Industry-funded audits and dashboards are useful, but we need independent measurements to ensure that AI search supports our digital ecosystem rather than disrupting it \cite{ingraphicdetail}.

\subsection*{Scientists}

Scientists have an obligation not just to study AI search, but to build the measurement infrastructure that shines a light on its impact on human judgment and the information economy. The core scientific risk is not that AI search is sometimes wrong (every information system is sometimes wrong). The risk is that it is persuasively, confidently, and selectively wrong, and that its errors are distributed unevenly across topics, populations, and institutions, while remaining hard to detect because the interface collapses uncertainty into a single fluent answer. Scientists' role is to quantify how often AI search substitutes for primary sources, how it changes belief formation and decision making, how it shifts traffic and incentives, and how its precision and accuracy fail in ways that matter.

First, scientists should build and maintain a shared, open methodology for measuring AI answer prevalence, citation patterns, and link-out behavior across platforms and countries. That means standardized instrumentation (browser-based audits, panel studies, and repeatable scraping protocols where lawful), agreed-upon topic taxonomies, and public reporting that is comparable over time. Crucially, these methods must assess whether cited sources actually contain the specific content being asserted. In practice, this requires claim-level support evaluation. Where platform terms constrain measurement, scientists should push for safe-harbor access to auditing interfaces and API endpoints designed explicitly for independent research.

Second, scientists should prioritize causal evidence over correlation. The most policy-relevant questions in this domain are counterfactual. For example, compared to traditional search, does AI search actually displace publisher traffic, increase overconfidence, reduce verification, accelerate decision-making at the expense of accuracy, or narrow the set of sources users engage with? These are questions for randomized controlled experiments and field studies. In laboratory settings, scientists can randomize interface features like claim-level citations, contested-claim badges, confidence highlighting, or source diversity, and measure changes in confidence, correctness, verification behavior (e.g. clicks, time on sources), and downstream decision quality (e.g., medical decisions, financial judgments, civic knowledge). In real-world settings, the gold standard will be partnerships that allow randomized experiments and A/B tests with privacy-preserving data collection. Where direct partnership is impossible, quasi-experimental designs like natural experiments, difference-in-differences, and regression discontinuities can produce credible estimates that regulators can consider when evaluating policy interventions. Without such studies it is difficult to make credible claims about the effects of AI search and therefore the policies needed to ensure healthy deployment and use of the technology.

Third, scientists should focus on the economics of information production. AI search is not just a new interface, it is potentially a new allocation mechanism for attention and revenue. Research should therefore measure how AI summaries change referral traffic, subscription conversions, and the distribution of attention across domains (news, reference, government, education, local publishers) and across the long tail. Scientists should build models that connect interface-level features like summary lengths, the number and placement of citations, and links to outcomes like publisher traffic, revenue, sustainability and content production and test these models with careful causal empirical analysis to understand their true effects. Information retrieval researchers, economists, computational social scientists, and legal scholars should work together on shared datasets and shared definitions of accuracy, substitution and complementarity.

Fourth, scientists should develop rigorous benchmarks for misinformation vulnerability and error modes. Red-teaming should be paired with prevalence measurement to understand how a model can fail, how often users are exposed to failure in the wild, and which populations, languages, or regions are the most vulnerable to such failures.

Finally, scientists should adopt norms of the highest standard including preregistration, shared material and code, and clear conflict-of-interest disclosures when research depends on platform access or funding. To maintain transparency, public-interest funding agencies, like the National Science Foundation and the National Institute of Standards and Technology, should treat independent AI search auditing as a core digital infrastructure and scientists who gain privileged access to platform data, through partnerships with OpenAI, Anthropic, or Google, should negotiate for publishable, generalizable results so that knowledge about societal impacts does not become a private asset held by the very systems being evaluated. If AI search is to become the default interface to knowledge, then independent measurement has to become the default interface to AI search.

\subsection*{Consumers}

Consumers also have an important role to play in ensuring the benefits and avoiding the pitfalls of AI search. Search engine literacy should expand to include AI search, how it works and how information access changes in an AI-first search regime. While we should teach our youth to be aware of and adept at using the unique features of AI search, we should all also be vigilant in educating ourselves on AI search. Such education should support behavioral practices that ensure access to a healthy information ecosystem and appropriate decision making in light of AI search answers. Some examples demonstrate this, although these examples are not exhaustive.

First, rules of thumb, like a ``two-click rule" for search used in consequential tasks could help ensure the vibrancy of information accessed in support of important decisions. For health, finance, legal or civic decisions, it is important to educate individuals and organizations to open at least two primary sources behind any AI answer. Institutions (schools, hospitals and courts) should codify this as policy. The speed dividend of AI search is real, but it should be paired with minimal verification habits \cite{effectsofllmbased}.

Second, we should learn to prefer sources with domain accountability. A “.gov” or ``.edu" URL is not a magic wand, but users should learn which domains and web pages are credible, authoritative and accountable and rely more on those for important information. We should also encourage users to privilege accountable domains when verifying facts found through other domains cited by AI search.

Finally, users should capture provenance in workflows. When an AI answer influences a decision (e.g. clinical notes, procurement memos, investigative drafts), we should require a citation list or screenshot of the answer state as well as a note that AI answers were consulted in making that decision. Such provenance notes appended to important workflows preserve a paper trail for later audits that could prove useful, for example, in judicial decisions or reviews of institutional practices.

\section*{Conclusion}

Search is the surface on which AI will affect human decision making on the greatest scale. But we know little about its prevalence or our exposure to AI search. The longitudinal evidence on the rise of AI search presented in this paper, which uses identical queries between 2024 and 2025 to control for behavioral explanations of AI search prevalence, demonstrates that platform policies are pushing AI search at a breakneck pace. Our research also shows that key policy choices about our information ecosystem, like the decision to turn AI search on for COVID queries, are being made without adequate transparency, and that AI search is fundamentally changing the variety, sources, credibility and political leanings of the information provided in search results. This is important because AI search will dramatically impact human decision making on a global scale and will affect our information ecosystem and the incentives for knowledge production that underpin our marketplace of ideas. The dramatic rise of AI search and its potential global impact necessitates more scientific inquiry and deliberate policy debates about how to harness this technology for its promise while avoiding its pitfalls. Toward that end, we are releasing all the data and code used in this paper, including the 2.8M search results and their information content, for replication and further research and analysis.\footnote{Data and code can be found hosted here: TBD.} We hope that this data and the ideas explored in this paper will spark a global debate about AI search, toward creating a brighter digital future.

\bibliography{refs_mnsc}
\bibliographystyle{sciencemag}

\clearpage

\setcounter{table}{0}
\renewcommand{\thetable}{S\arabic{table}}
\setcounter{figure}{0}
\renewcommand{\thefigure}{S\arabic{figure}}

\begin{center}
\section*{Supplementary Materials for\\ The Rise of AI Search: Implications for Information Markets and Human Judgment at Scale}

Sinan Aral$^{1\ast}$,
Haiwen Li$^{1}$,
Rui Zuo$^{1}$\\ 

$^{1}$Massachusetts Institute of Technology, Cambridge, MA, USA.\\
\small$^\ast$Corresponding author. Email: sinan@mit.edu\\

\end{center}

\subsubsection*{This PDF file includes:}
Materials and Methods\\
Robustness Checks\\
Figures S1 to S5\\
Tables S1 to S2\\

\newpage


\subsection*{Materials and Methods}
To examine global exposure to AI search, we measured the prevalence of AI Overview responses on Google Search and how this prevalence varies by country, query topic, and query style. Our analysis involved three waves of data collection across 2024 and 2025. The first wave of data collection occurred from October 8 to November 8, 2024 and included the seven countries in which AI Overviews were publicly available in 2024: the United States, United Kingdom, India, Mexico, Brazil, Japan, and Indonesia. The second wave occurred from August 28 to September 5, 2025 and covered the same seven countries. The third wave occurred from October 14 to November 13, 2025, after AIO began expanding internationally, and during which we collected data from an additional 236 countries, for a total of 243 countries.

\subsubsection*{Query Selection}
We collected a total of 11,372 real-world user search queries which were randomly sampled from the following data sources, with all non-English language queries excluded:
\begin{itemize}
    \item 3561 queries from Google’s Natural Questions\\(https://ai.google.com/research/NaturalQuestions).
    \item 3386 queries from Microsoft’s MSMARCO\\(https://microsoft.github.io/msmarco/).
    \item 3130 queries from Quora Question Pairs\\(https://quoradata.quora.com/First-Quora-Dataset-Release-Question-Pairs).
    \item 428 queries from 2015-2020 Google Trends keywords (US and Global)\\(https://trends.google.com/trends/). 
    \item 283 queries from most-searched Google queries (US and Global)\\(https://www.semrush.com/blog/most-searched-keywords-google/).
    \item 200 coding queries from Microsoft’s Search4Code\\(https://github.com/microsoft/Search4Code).
    \item 168 queries from a COVID-related search query dataset\\(https://figshare.com/articles/dataset/Search\_query\_lists/12398309).
    \item 116 queries from a product search query dataset\\(https://huggingface.co/datasets/trec-product-search/product-search-2024-queries).
    \item 100 queries from Amazon shopping query dataset\\(https://github.com/amazon-science/esci-data).
\end{itemize}

\subsubsection*{Query Labeling}
Table~\ref{si-tab:query-style} and Table~\ref{si-tab:query-topic} summarize the descriptive statistics of the sampled queries by style and topic. We used LLMs for query labeling, given their demonstrated capacity to understand natural language and their growing adoption in both research and industry settings for scalable and consistent data annotation \cite{tan2024largelanguagemodelsdata, annotationpnas}. 

Specifically, we used GPT-4o-mini to annotate each query into one of three styles: question, statement, or navigational. The labeling prompt was:
\begin{quote}
Identify the format of the following search engine query: whether it is a statement, an interrogative query (question), or a navigational query (an internet search with the clear intent of finding a specific website or web page; a navigational query can usually be the name of a brand, platform, organization, or specific URL, intending to navigate directly to that destination).
\end{quote}

For topic labeling, we followed a multi-step procedure. First, we used GPT-4o-mini to assign an initial topic label to a random subset of queries with the following prompt: 
\begin{quote}
    Label the following search engine query with one topic based on the subject matter.
\end{quote}
We next manually selected and grouped the initial labels to build a topic list. We then prompted GPT-4o-mini to label the remaining queries by selecting a single topic from this list, using the following prompt:
\begin{quote}
Label the following search engine query with one topic based on the subject matter. Your answer should be a topic label from the provided list. If the query doesn't fit in to any of the given topics, come up with your own topic.
\end{quote}
Finally, we further grouped the labels into seven categories: ``General Knowledge'', ``Health'', ``Internet, Technology, Media'', ``Shopping'', ``Lifestyle'', ``Business, Finance, Employment'' and ``Covid''. The ``Covid'' category was specifically defined to include all queries mentioning ``Covid'' or ``Coronavirus.''

\subsubsection*{Search Result Collection}
Search results were collected using SerpApi's Google Search Engine Results API (https://serpapi.com). To simulate searches originating from different countries, we set the API request's location parameter accordingly. If a Google search returned an AI Overview, the API output contained a corresponding information block (the search result). We recorded the existence of the AI Overview response and scraped the content of the information block for further analysis. We did the same for traditional search results and featured snippets, for comparison. In total, this process yielded 2,843,000 search results ($11372 * 7+11372 *243$).

\subsubsection*{Predicting the Appearance of AI Overview Responses}
We tested the predictive power of three feature sets—country, query style, and query topic—in predicting whether a search returned an AI Overview. The task was framed as binary classification, with ``0'' denoting no AI Overview returned and ``1'' denoting the presence of one.

We trained Logistic Regression models using each feature set independently, as well as a full model incorporating all three. We chose Logistic Regression for its ability to handle categorical data through one-shot encoding, its efficiency in large binary classification tasks, and its provision of interpretable coefficients that can be converted to odds ratios.

Model performance was assessed using the Area Under the Receiver Operating Characteristic Curve (AUC), which captures the trade-off between true and false positives and is independent of classification thresholds. To obtain robust estimates, we applied 5-fold cross-validation repeated 100 times, generating 500 AUC scores per model. This approach allowed direct comparison of feature sets using the same model framework and provided insight into which features most strongly predict AI Overview appearances.

To further analyze feature importance, we fitted the full model with all feature sets and calculated odds ratios from the model coefficients. The odds ratio for each feature reflects the multiplicative change in the odds of an AI Overview appearing associated with that feature. Values greater than 1 indicate increased likelihood, while values less than 1 indicate decreased likelihood. Figures 2A and 2B in the main text show the odds ratios for predictive features and compare the performance of the Logistic Regression models by AUC.

\subsubsection*{Response Variety}

To measure the dispersion of information in AI Overview and traditional results, we first convert each element into a sentence embedding using SBERT, a widely adopted benchmark for sentence-level semantic representations \cite{reimers-2019-sentence-bert}. We then apply the information uniqueness metric adapted from Aral and Dhillon \cite{aral2023exactly}, which quantifies dispersion by calculating the average pairwise dissimilarity between all elements within a result unit. We chose embedding-based information uniqueness to measure diversity for several methodological reasons. First, embeddings capture semantic meaning in continuous vector space, allowing us to quantify how distinct pieces of information are regardless of surface-level differences in vocabulary or sentence structure. This continuous representation avoids the sensitivity to arbitrary categorization inherent in topic modeling approaches, where diversity measurements depend heavily on predefined topic boundaries that may not align well with the scope of search results. Second, information uniqueness is easily interpreted. It measures the average semantic distance between each element and all others within a result unit, directly quantifying the degree to which users encounter new information as they read through a search result. Higher values indicate that each individual element is more semantically distinct from one another, reflecting greater diversity.

\begin{figure}[h!]
    \centering
    \includegraphics[width=0.8\linewidth]{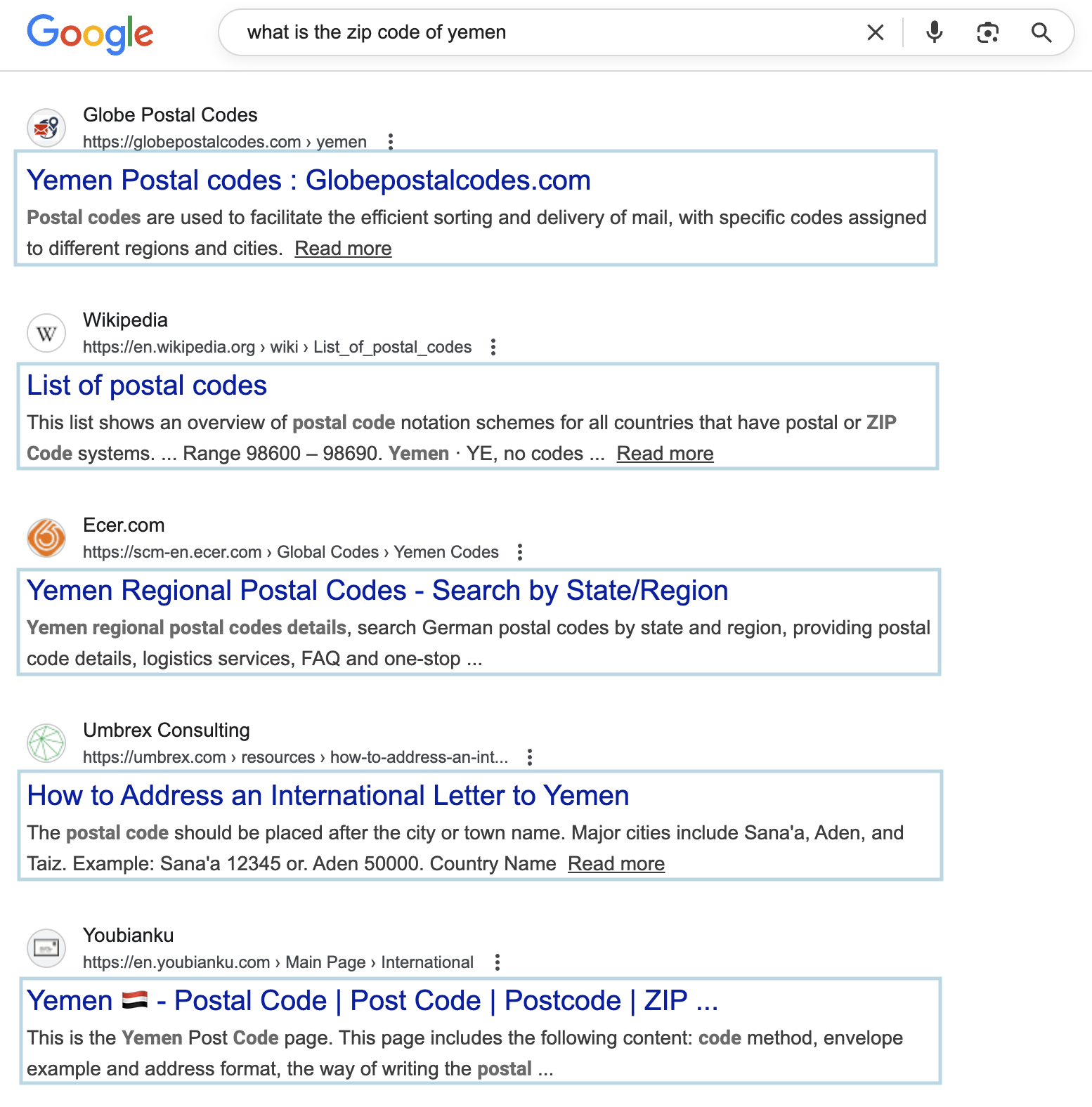}
    \caption{Traditional Results Unit of Analysis (in Light Blue Boxes)}
    \label{fig:org_embedding_unit}
\end{figure}

\begin{figure}[h!]
    \centering
    \includegraphics[width=0.8\linewidth]{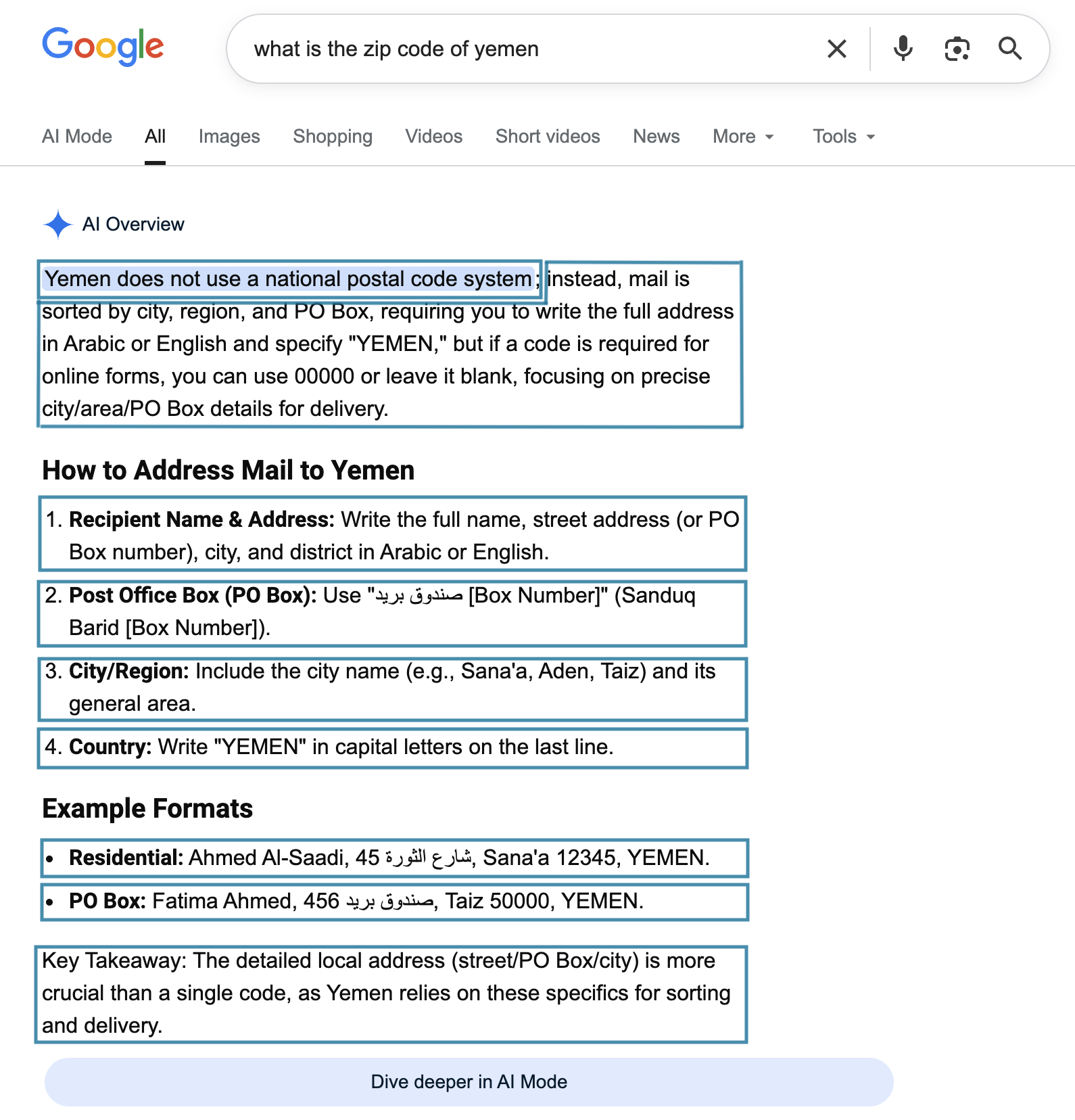}
    \caption{AI Overview Unit of Analysis (in Dark Blue Boxes)}
    \label{fig:aio_embedding_unit}
\end{figure}

\paragraph{Unit of Analysis} A single AI Overview or the first page of traditional search results is denoted as unit $i$, with each constituent element within that unit denoted as $ij$. For traditional results, each unit $i$ comprises the full first page of search results, and each element $ij$ corresponds to the text snippet displayed for each result, as illustrated in the light blue boxes in Figure~\ref{fig:org_embedding_unit}. For AI Overviews, each unit $i$ is the complete AI-generated response, and each element $ij$ is either an individual sentence (when the response is displayed as paragraphs) or an individual bullet point (when the response is displayed in list formatting), as illustrated in the dark blue boxes in Figure~\ref{fig:aio_embedding_unit}.

\paragraph{Calculating Response Variety} For each unit $i$, response variety, measured by information uniqueness is calculated as:
$$
\text{InformationUniqueness}_{i} =\frac{1}{n_i}\sum_{j=1}^{n_i} \frac{1}{n_i-1}\sum_{j'\neq j}^{n_i}\left[1-\cos(\Gamma_{ij}, \Gamma_{ij'})\right]
$$
where $\Gamma_{ij}$ represents the SBERT embedding of element $ij$, $n_i$ is the number of elements in unit $i$, and $1-\cos(\Gamma_{ij}, \Gamma_{ij'})$ measures the cosine distance between each pair of embeddings. The metric averages these pairwise distances across all elements. Higher values indicate the elements are more semantically distinct and therefore. We calculated information uniqueness separately for AI Overview and traditional results within each country-query pair, as shown in Figure 2C.

\subsubsection*{Domain Traffic Ranking Distributions}
We measured the popularity of reference URLs and links in AI Overviews and traditional search results to understand the extent to which AI and traditional search linked to more popular or more long tail websites. We measured domain popularity using global traffic rankings rather than exact traffic volumes, as exact domain-level traffic data is difficult to obtain reliably. Domain rankings were obtained from the Cisco Umbrella Top 1 Million list, which ranks domains by relative Domain Name System query frequency across Cisco's global network infrastructure handling over 100 billion requests daily from 65 million active users in more than 165 countries \cite{cisco_umbrella}. Each domain appearing in the search results was matched against this list, achieving coverage of 78.8\% for traditional results and 76.7\% for AI Overview references. We categorize the domains into three tiers based on their ranking: Top 1k (high-traffic, widely recognized domains), Top 1K–1M (moderate-traffic domains), and beyond Top 1M (unmatched domains, indicating low traffic or niche presence). The 21.2\% of traditional domains and 23.3\% of AI Overview reference domains that did not appear in the Cisco list were assigned to this third category. For each country- query unit with both AI Overview Reference and Traditional Results, we compute the fraction of domains falling into each tier separately, for each result type, as presented in Figure 2D.

\subsubsection*{Domain Credibility and Political Leaning}
Domain credibility and political leaning ratings were obtained from Media Bias/Fact Check (MBFC) (\url{https://mediabiasfactcheck.com/mbfc-ratings-by-the-numbers/}), a widely used third-party fact-checking resource frequently referenced in academic research on misinformation and media quality \cite{mbfc}. The MBFC database contains ratings for over 9,200 actively monitored media sources, evaluating each along two dimensions: factual reporting and political bias. Factual reporting ratings range from Very High to Very Low, derived from a weighted composite of failed fact checks (40\%), sourcing quality (25\%), transparency (25\%), and one-sidedness (10\%). Political bias scores range from -10 (extreme left) to +10 (extreme right), calculated from assessments of economic policy positions (35\%), social values (35\%), straight news reporting balance (15\%), and editorial bias (15\%). These ratings were matched to all domains appearing in the search results, achieving coverage of 20.8\% for traditional results and 18.7\% for AI Overview references. The matched ratings were used to construct both Figure 2E (source credibility) and Figure 2F (source political leaning).

\paragraph{Domain Credibility}
MBFC classifies domain credibility into three tiers: high credibility (sources deemed trustworthy for reliable information), medium credibility (sources inconsistent in factual accuracy or lacking transparency), and low credibility (questionable sources or those promoting conspiracy theories and pseudoscience). For clarity, we combine medium and low credibility domains into a single ``low credibility" category. For each country–query unit containing both AI Overview references and traditional results, we calculated the share of domains in both credibility tiers separately, for AI and traditional search results, as presented in Figure 2E.

\paragraph{Domain Political Leaning}

MBFC classifies domain political stance into nine categories: Right, Right-Center, Least Biased, Pro-Science, Left-Center, Left, Questionable, Conspiracy-Pseudoscience, and Satire. For clarity, we consolidated these into four groups: right (Right and Right-Center), center (Least Biased and Pro-Science), left (Left-Center and Left), and unclear (Questionable, Conspiracy-Pseudoscience, and Satire). For each country-query unit with both AI Overview Reference and Traditional Results, we compared the average share of domains classified as left-leaning, center, and right-leaning separately for each type of results, as presented in Figure 2F. We excluded the ``Questionable, Conspiracy-Pseudoscience, and Satire" category because it conflated known satirical sources like \textit{The Onion} with misinformation and conspiracy theory sources and because it did not clearly delineate right- left- or center-oriented information. This category comprised less than .5\% of the data.

\subsection*{Robustness Checks}

\subsubsection*{Writing Style}
Because AI Overviews are written in a single voice while traditional results combine snippets from many authors, lower response variety in AI Overviews could reflect stylistic uniformity rather than information content. To separate the two, we paraphrased every element in a stratified 25\% random sample of queries (strata defined by country, topic and style; strata with fewer than 40 queries were included in full) using DIPPER \cite{krishna2023paraphrasing}, a T5-XXL paraphrase model with controllable lexical and order diversity. We set both lexical and order diversity to 80 (on a 0--100 scale), applied identical settings to AI Overviews and traditional results, and recomputed information uniqueness on the paraphrased elements. After paraphrasing, AI search results continue to exhibit lower response variety than traditional results in every category, and the overall gap is larger than with the original text (Fig. S3; compare Fig. 2C).

\subsubsection*{Embedding Model}
We re-encoded all elements with Harrier (harrier-oss-v1-0.6b) \cite{harrier2026}, a multilingual embedding model from Microsoft that reported state-of-the-art results on the Multilingual MTEB v2 benchmark at release. Harrier differs from the SBERT model used in the main analysis (all-MiniLM-L6-v2, an encoder-only model producing 384-dimensional embeddings by mean pooling) in both architecture and scale: it is a 0.6B-parameter decoder-only model producing 1,024-dimensional embeddings by last-token pooling. Elements from both AI Overviews and traditional results were encoded without a task instruction. With Harrier embeddings, AI search results again exhibit lower response variety than traditional results in every category (Fig. S4). Absolute values are not comparable across embedding models; only the difference between AI and traditional results within each model is informative.

\subsubsection*{Query Labels}
We re-labeled all 11,372 queries for style and topic with two additional LLMs, GPT-5.6 Terra (\texttt{gpt-5.6-terra}) and Claude Sonnet 4.6 (\texttt{claude-sonnet-4-6}), using the prompts above. The results are broadly similar under alternative labeling regimes (Fig. S5).

\clearpage

\newpage

\newpage
\begin{figure}[h!]
    \centering
    \includegraphics[width=0.8\linewidth]{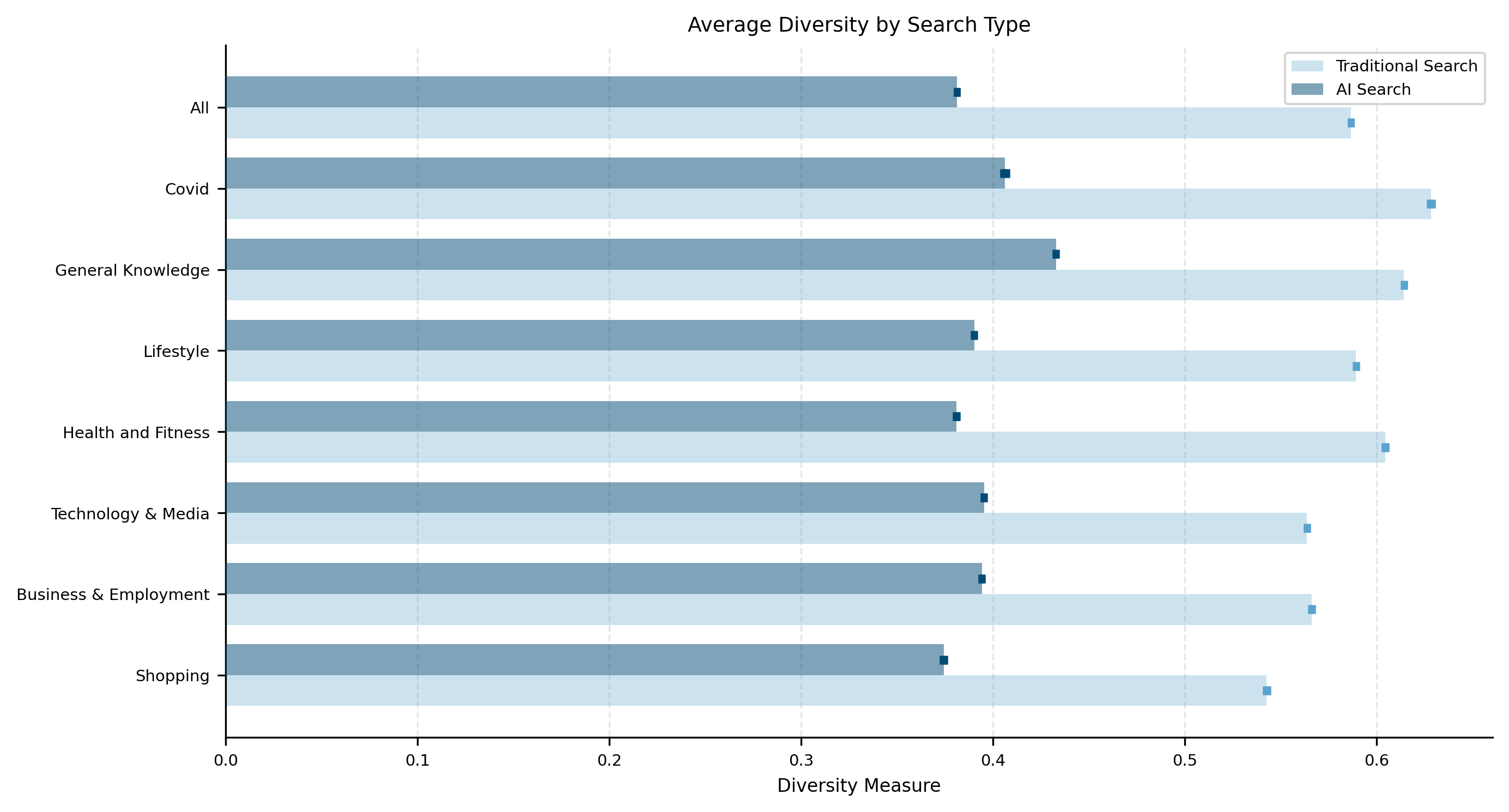}
    \caption{\textbf{Response Variety after Paraphrasing.} Average response variety (information uniqueness) of traditional and AI search results, by topic and for all queries, after paraphrasing every element in a stratified 25\% random sample with DIPPER (lexical and order diversity of 80). Compare Fig. 2C.}
    \label{fig:robust_style}
\end{figure}

\newpage
\begin{figure}[h!]
    \centering
    \includegraphics[width=0.8\linewidth]{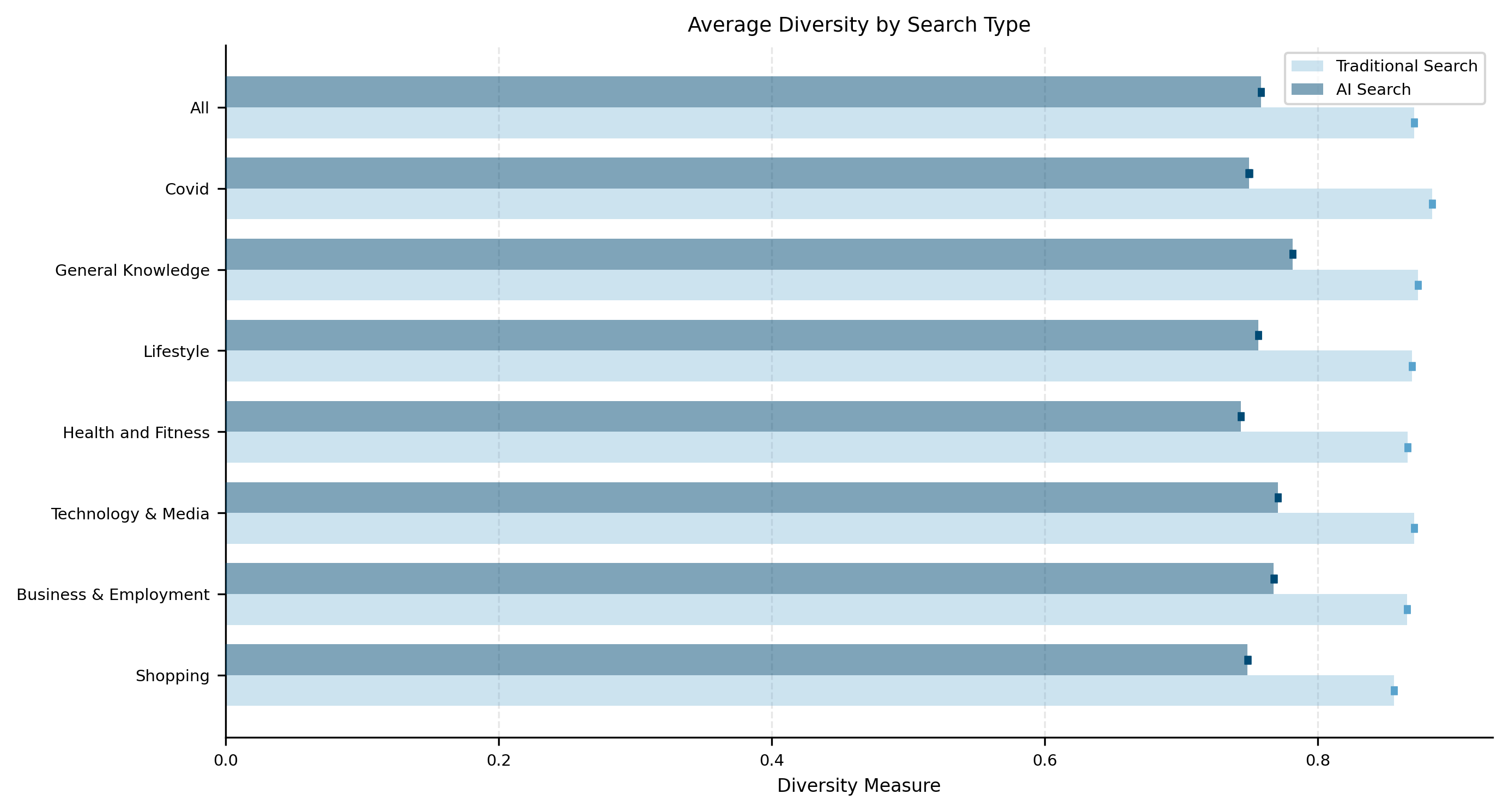}
    \caption{\textbf{Response Variety with Harrier Embeddings.} Average response variety (information uniqueness) of traditional and AI search results, by topic and for all queries, computed from Harrier (harrier-oss-v1-0.6b) embeddings. Absolute values are not comparable with the SBERT-based values in Fig. 2C.}
    \label{fig:robust_embedding}
\end{figure}

\newpage
\begin{figure}[h!]
    \centering
    \begin{overpic}[width=0.62\linewidth]{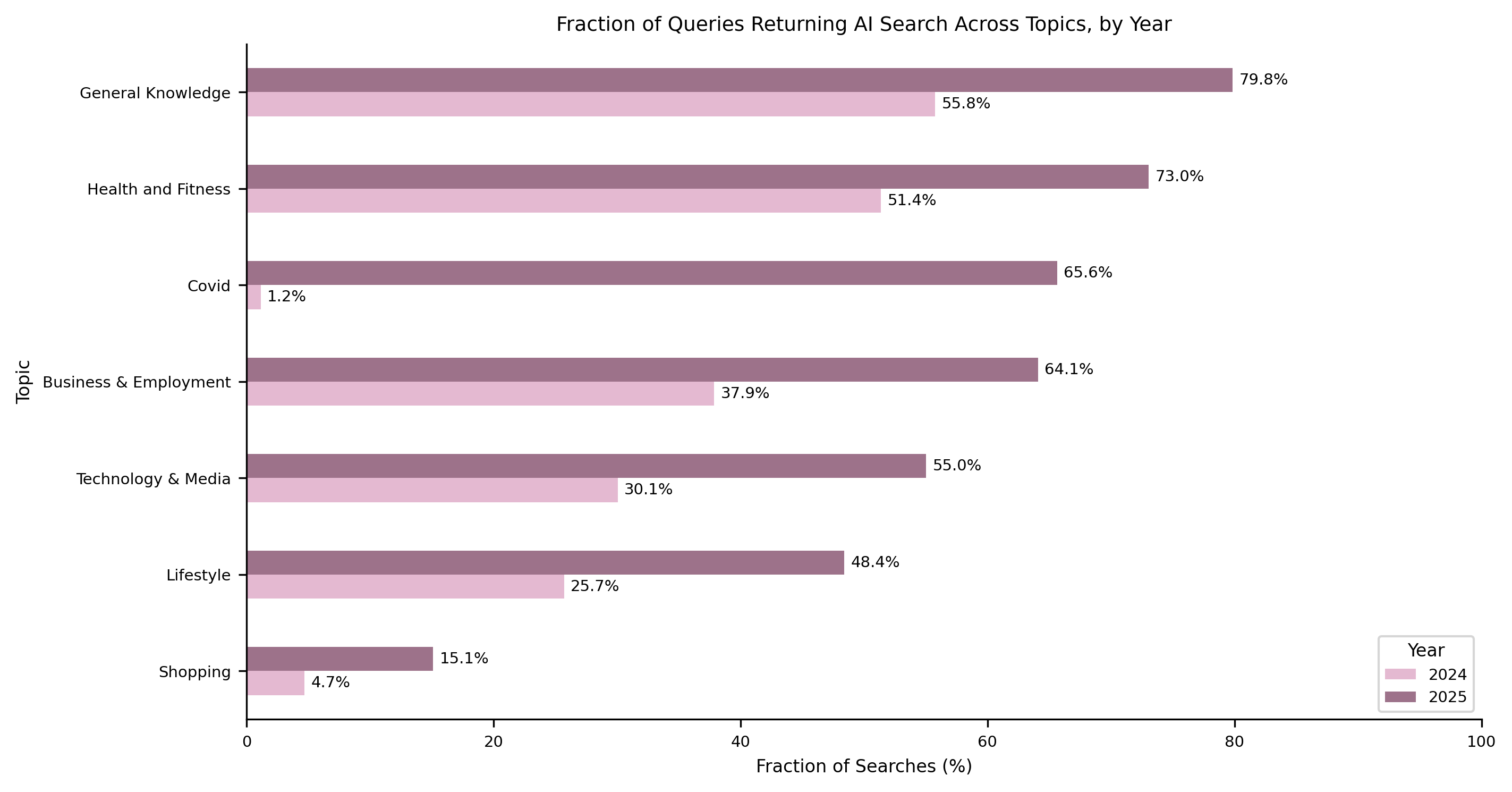}\put(1,49){\textbf{A}}\end{overpic}

    \medskip
    \begin{overpic}[width=0.62\linewidth]{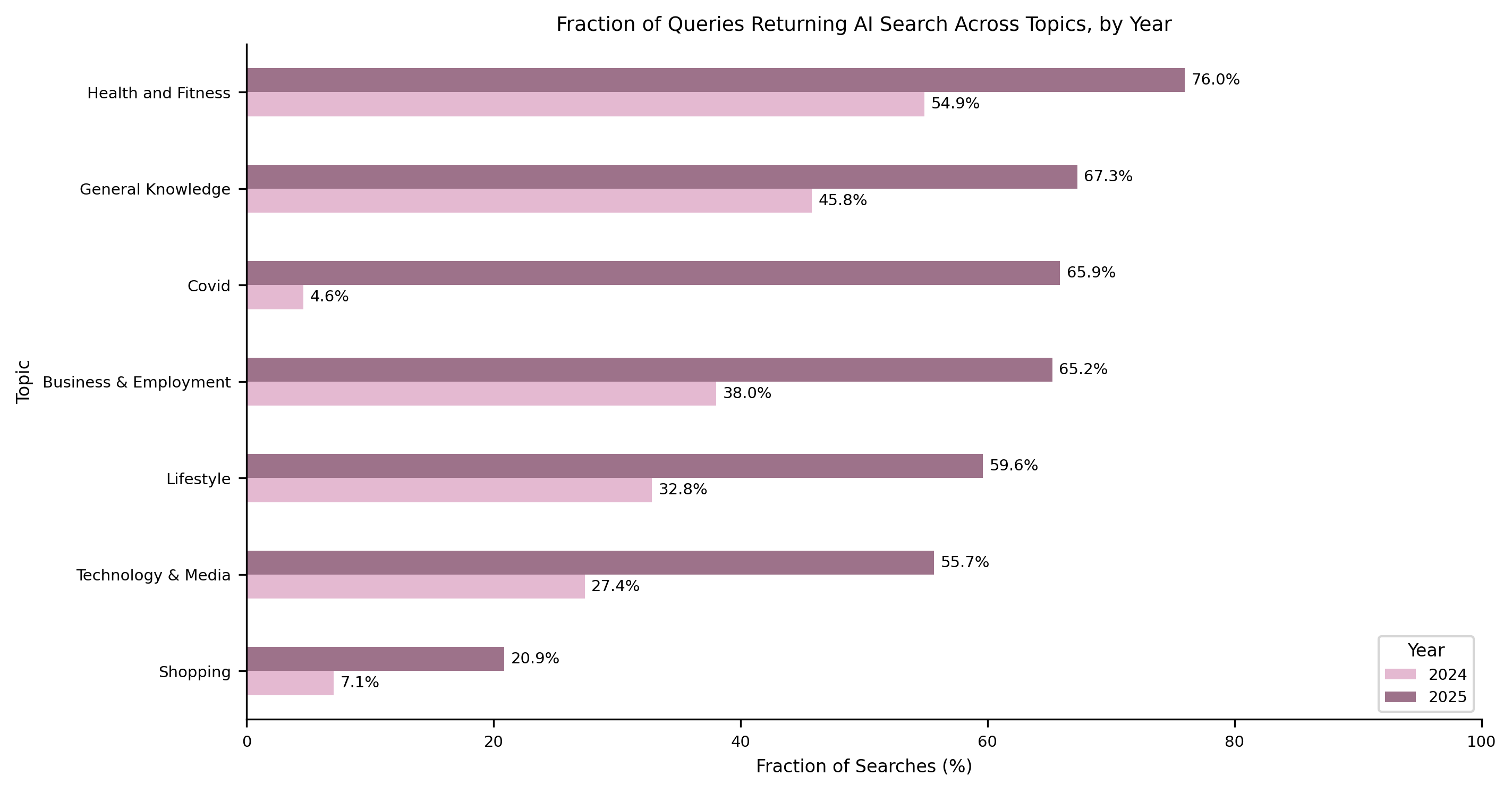}\put(1,49){\textbf{B}}\end{overpic}

    \medskip
    \begin{overpic}[width=0.62\linewidth]{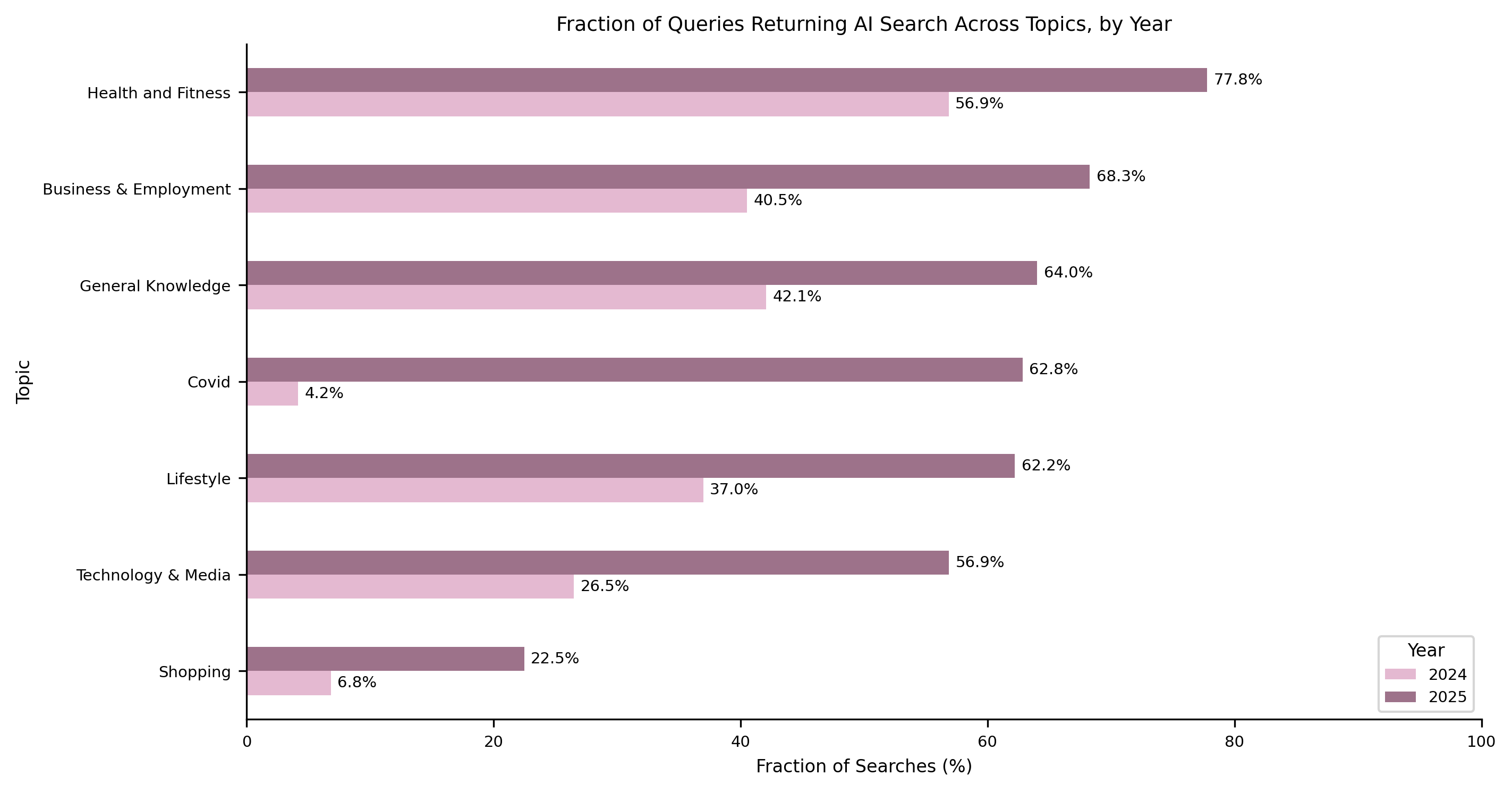}\put(1,49){\textbf{C}}\end{overpic}
    \caption{\textbf{Fraction of Queries Returning AI Search Results across Topics under Alternative Query Labels.} Fraction of queries returning AI search results by topic in 2024 and 2025, in the 7 countries exposed to AI search in 2024, with topic labels from (\textbf{A}) GPT-4o-mini (original labels, as in Fig. 1F), (\textbf{B}) GPT-5.6 Terra, and (\textbf{C}) Claude Sonnet 4.6.}
    \label{fig:robust_labels}
\end{figure}

\newpage
\begin{table}[h!]
    \centering
    \caption{\textbf{Search Query Styles}}
    \label{si-tab:query-style}
    \begin{tabular}{lr}
    \\
    \hline
    Style & count \\ 
    \hline
    Question & 7560 \\ 
     Statement & 2360 \\ 
    Navigational & 1452 \\ 
    \hline
\end{tabular}
\end{table}

\begin{table}[h!]
\centering
\caption{\textbf{Search Query Topics}}
\label{si-tab:query-topic}
\begin{tabular}{lr}
\\
 \hline
Topic & Count \\ 
  \hline
Internet, Technology, Media & 3421 \\ 
General Knowledge & 2467 \\ 
Business, Finance,Employment & 2137\\ 
 Lifestyle & 2017 \\ 
Health & 808 \\ 
 Shopping & 448 \\ 
 Covid & 74 \\ 
   \hline
\end{tabular}
\end{table}

\end{document}